\documentstyle[twoside,fleqn,espcrc2,epsfig]{article}

\def\spose#1{\hbox to 0pt{#1\hss}}
\def\ltapprox{\mathrel{\spose{\lower 3pt\hbox{$\mathchar"218$}}
 \raise 2.0pt\hbox{$\mathchar"13C$}}}
\def\gtapprox{\mathrel{\spose{\lower 3pt\hbox{$\mathchar"218$}}
 \raise 2.0pt\hbox{$\mathchar"13E$}}}
\def\inapprox{\mathrel{\spose{\lower 3pt\hbox{$\mathchar"218$}}
 \raise 2.0pt\hbox{$\mathchar"232$}}}

\newcommand{\de}{\partial}

\newcommand{\crit}{\mbox{\scriptsize crit}}

\newcommand{\sea}{\mbox{\scriptsize sea}}

\newcommand{\val}{\mbox{\scriptsize val}}
\newcommand{\PCAC}{\mbox{\scriptsize PCAC}}

\newcommand{\SW}{\mbox{\scriptsize SW}}
\newcommand{\Wilson}{\mbox{\scriptsize Wilson}}

\newcommand{\etal}{{\em et al.}}

\newcommand{\NPB}{Nucl.Phys.B}

\newcommand{\<}{\langle}
\renewcommand{\>}{\rangle}

\newcommand{\be}{\begin{equation}}
\newcommand{\ee}{\end{equation}}
\newcommand{\bea}{\begin{eqnarray}}
\newcommand{\eea}{\end{eqnarray}}

\title{A First Taste of Dynamical Fermions 
with an $O(a)$ Improved Action.}

\author{
M.~Talevi\address{
Department of Physics \&\ Astronomy, University of Edinburgh,
The King's Buildings, EH9 3JZ (UK)}
\thanks{Talk presented at Lattice '97, XV International Symposium on Lattice Field Theory, Edinburgh (UK), July 1997.}, 
{\em for the UKQCD Collaboration}
}

\begin{document}

\begin{abstract}
We present the first results obtained by the UKQCD Collaboration using a
non-perturbatively $O(a)$ improved Wilson quark action with 
two degenerate dynamical flavours. 
\end{abstract}

\maketitle

The Symanzik improvement program
is an attractive method which allows reduction of discretization errors 
order by order in $a$ in physical quantities.  
Recently, the improvement counterterm $c_{\SW}(g_0^2)$ for the improved 
action \cite{SW}
\begin{equation}
S=S_{\Wilson}+
ac_{\SW}\int dx \bar\psi\frac{i}{4}\sigma_{\mu\nu}F_{\mu\nu}\psi
\label{eq:SW}
\end{equation}
has been computed non-perturbatively, thus yielding full 
$O(a)$ improvement, both in the quenched approximation \cite{Luscher}, 
and with $N_f=2$ dynamical fermions \cite{Jansen}.  

The UKQCD Collaboration intends to exploit the action (\ref{eq:SW})
with two degenerate sea quarks for its Dynamical Fermions Project.   
The calculations have been carried out on a Cray T3E recently installed in 
Edinburgh.  The version on which these results were obtained consisted of 
96 processor elements (PE), each of 900 MFlops peak speed, and our code
sustained a speed of 25-30 GFlops.

The algorithm that has been implemented is a Generalized Hybrid Monte Carlo 
(GHMC) with even-odd preconditioning.  A variant of the 
Sexton-Weingarten integration scheme has been employed.  We refer the reader
to \cite{Sroczynski} for all details
of the implementation, optimization and testing of the algorithm. 

The simulation parameters are summarized in tab.~\ref{tab:params}.  
The value of $c_{\SW}$ used was kindly comunicated to us by the 
{\sc Alpha} Collaboration \cite{Jansen},
albeit at a very preliminary stage.
Hence the discrepancy between the number we chose with respect 
to the final result presented in \cite{Jansen}.  
Since this is the first attempt to do hadronic
physics with such an improved action in the dynamical case, even with 
slightly incomplete improvement we hope 
to gain much insight in terms of both the performance 
of the code and of the choice of the parameters for future runs.
Thus, the results presented here should be regarded primarily as a testing 
ground to explore the feasibility of the new action and algorithm.
The choice of a large coupling was also motivated on such grounds.  
Given the exploratory character of the simulations, we have limited ourselved
to using local sources and have not exploited any form of smearing.  
Nevertheless, some light spectrum physics in the mass region around the
strange quark mass can still be extracted.

\begin{table}[t]
\begin{tabular}{rrrrr}
\hline
$\beta$ & $c_{\SW}$ & $L^3\cdot T$ & $\kappa_{\sea}$  & Conf \\ \hline \hline
 5.2 &  1.76 & $12^3\cdot 24$ &  $0.1370$  & 50 \\
     &       &                &  $0.1380$  & 50 \\
     &       &                &  $0.1390$  & 50 \\
     &       &                &  $0.1395$  & 50 \\ \hline 
\end{tabular}
\caption{Simulation parameters.}
\vspace{-1.0truecm}
\label{tab:params}
\end{table}

For each of the four values of $\kappa_{\sea}$ we have simulated valence quarks
with the same set of four $\kappa$'s.  One must keep in mind that in order to
take a sensible chiral limit, we need to put 
$\kappa\equiv\kappa_{\sea}=\kappa_{\val}$,
but the full $(\kappa_{\sea},\kappa_{\val})$ plane can be exploited for
``strange'' physics \cite{Guesken}. At $\beta=5.2$ the $\pi/\rho$ mass ratios
for the relevant $\kappa$'s are shown in tab.~\ref{tab:masses}.
These numbers show how relatively far we are from the chiral region.

\section{Setting the scale}

The critical point was determined from a linear extrapolation in $1/\kappa$
to vanishing $M_{\pi}^2$.  The linear ansatz is well supported by the 
numerical data, cf.~tab.~\ref{tab:masses}.  
The result $\kappa_{\crit}=0.14033(3)$ was confirmed
by the scaling behaviour of the GHMC algorithm as a function of 
$\kappa_{\sea}$ \cite{Sroczynski}.  

\begin{figure}[t]
\vspace{-2.5truecm}
\begin{center}
\setlength{\epsfxsize}{10cm}
\setlength{\epsfysize}{10cm}
\epsfig{figure=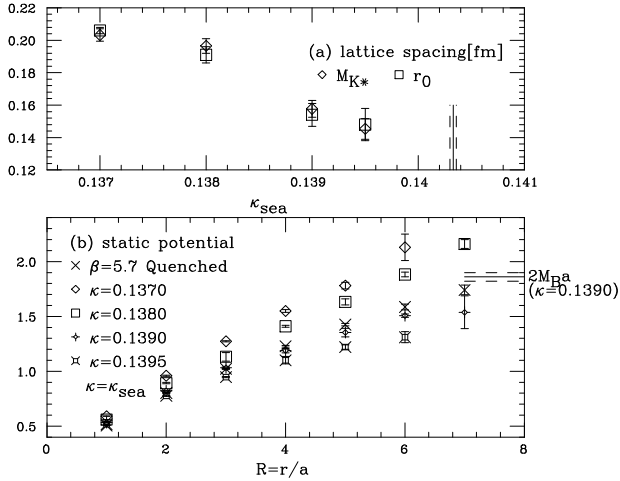,height=11cm,angle=0}
\vspace{-3.5truecm}
\caption{(a) Lattice spacing and
(b) Static quark potential as a function of the sea quark mass.}
\end{center}
\vspace{-1.0truecm}
\label{fig:latspac}
\end{figure}

The lattice spacing was determined in two independent ways, 
cf.~fig.~\ref{fig:latspac}a:  from the scale
$r_0$ \cite{Sommer} and from $M_{K^*}$ \cite{Allton}.  We stress that neither
method requires any extrapolation to $\kappa_{\crit}$ and 
are completely consistent at all values of $\kappa_{\sea}$.  We view the
standard approach of fixing the scale from the chiral extrapolation of
$M_{\rho}$ as unreliable, given the strong non-linear behaviour in $1/\kappa$ 
shown by the data, cf.~tab.~\ref{tab:masses}.  
Moreover, in full QCD the $\rho$ can decay and it
is therefore dubious to extrapolate to a region where it becomes 
unstable.  In fig.~\ref{fig:latspac}b we show the static potential from which
$r_0$ was calculated, together with the value of twice the static 
heavy-light meson mass, calculated for $\kappa=0.1390$, 
which is the upper limit for the potential in the case of string breaking.  
Plotting the potential in units of $r_0$, we recover a universal curve,
as shown in \cite{Guesken}.  Our data does not exclude string breaking
although the distances $r/r_0$ are too small and  we need to explore larger 
distances to be able to draw any definite conclusion.  

The strong dependence of $a$ on 
$\kappa_{\sea}$ indicates the non-negligible effect of the improvement 
counterterm of the coupling, necessary for the full $O(a)$ improvement of the 
action
\begin{equation}
\tilde g_0^2=g_0^2(1+b_g(g_0^2)\,m_qa).
\end{equation}
To one-loop in perturbation theory $b_g(g_0^2)=0.012\,N_f\,g_0^2$ \cite{Sint},
 which shows why this counterterm is not needed in the quenched approximation 
$(N_f=0)$.  A non-perturbative method to evaluate $b_g$ has been proposed
in \cite{Martinelli}, and its application is currently under investigation by 
us.  Preliminary results at higher values of $\beta$ indicate a much less 
pronounced mass dependence of the lattice spacing, as expected.

\section{Light Spectrum}

\begin{table}[t]
\begin{tabular}{ccccc}
\hline
$\kappa$  & $M_{\pi}^2a$ & $M_{\pi}/M_{\rho}$ & $M_N/M_{\rho}$ & $J$ \\ 
\hline \hline
$0.1370$  &  1.21(1)   & 0.855(4) & 1.56(2) & 0.356(4) \\
$0.1380$  &  0.87(1)   & 0.824(5) & 1.51(2) & 0.343(8)\\
$0.1390$  &  0.48(1)   & 0.787(10)& 1.56(2) & 0.367(7)\\
$0.1395$  &  0.31(1)   & 0.738(16)& 1.59(4) & 0.371(9)\\ \hline
\end{tabular}
\caption{Some light hadron results for different $\kappa$'s.}
\vspace{-0.8truecm}
\label{tab:masses}
\end{table}

\begin{figure}[t]
\vspace{-2.5truecm}
\begin{center}
\setlength{\epsfxsize}{10cm}
\setlength{\epsfysize}{10cm}
\epsfig{figure=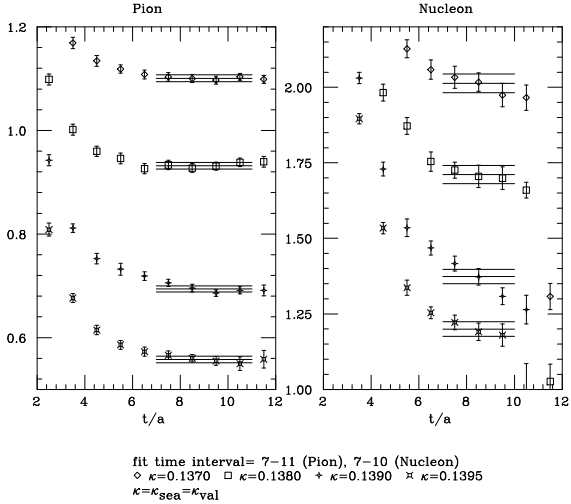,height=11cm,angle=0}
\vspace{-3.5truecm}
\caption{Effective mass of the pion and nucleon 
for different $\kappa$'s.}
\end{center}
\vspace{-0.8truecm}
\label{fig:m_eff}
\end{figure}

We now give some preliminary results of the calculation 
of the light hadron spectrum.  
The standard way to address this point is to show the
effective mass.  In fig.~\ref{fig:m_eff} 
we report it for the pion and for the nucleon, for the different $\kappa$'s.
The lines convey some of the information from the exponential fit:
the central line is the value of the mass, the outer lines denote the error
spread, and the extension of the lines the fit interval.
The error has been obtained with the jacknife method.
The information we can gather from them is that the plateau is quite evident
for the pion, even with a relatively small volume, 
and the fit is quite reliable (a similar situation occurs for the $\rho$),  
taking into consideration the fact that this result
is obtained without any type of smearing. On the other hand, and not 
surprisingly, the situation is worse for the nucleon.  In both cases we 
expect a better isolation of the ground state from smearing.  A full
statistics study, including the effects of smearing, is underway 
\cite{UKQCD}.  Some of our results are summarized in 
tab.~\ref{tab:masses}, in which we also present the variable 
$J=M_{K^*}dM_{\rho}/dM_{\pi}^2$. 

\section{Improvement of the Axial Current}

\begin{figure}[t]
\vspace{-2.0truecm}
\begin{center}
\setlength{\epsfxsize}{10cm}
\setlength{\epsfysize}{10cm}
\epsfig{figure=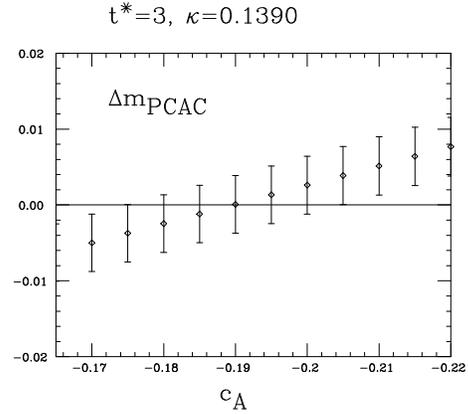,height=10cm,angle=0}
\vspace{-3.5truecm}
\caption{$\Delta m_{\PCAC}$ as a function of $c_A$.}
\end{center}
\vspace{-0.8truecm}
\label{fig:dm_PCAC}
\end{figure}

The improved on-shell local axial current 
\begin{equation}
A^I_{\mu}(x,c_A)=A_{\mu}(x)+ac_A\de_{\mu}P(x)
\end{equation}
can be determined by enforcing PCAC, as done in the Schr\"odinger Functional
formalism \cite{Luscher}.  
An equivalent approach, proposed in \cite{Martinelli}, can
be applied using standard hadronic correlators.  The current 
quark mass (averaged over space)
\begin{equation}
m_{\PCAC}(t,c_A)=\frac{1}{2}\frac{\<\de_tA^I_0(t,c_A)P(0)\>}{\<P(t)P(0)\>}
\end{equation}
is required to be independent of $t$ (up to $O(a^2)$) if the improvement 
coefficients $c_{\SW},c_A$ and $b_g$ are correctly fixed.  
We know that this is not completely the case for the present simulation, 
but we are here interested mainly in showing the viability of the method.  
Plotting 
$\Delta m_{\PCAC}(t^*,c_A)=m_{\PCAC}(T/2,c_A)-m_{\PCAC}(t^*,c_A)$,
we fix $c_A$ by looking at the zero intercept.  In fig.~\ref{fig:dm_PCAC}
we report a typical example, which shows that, for a given value of $t^*$,
there is a distinct cross-over from negative to positive values at a fixed
value of $c_A$, which can be precisely determined by a linear interpolation. 
We have checked that other values of $t^*$ do not yield
contradictory information, as $\Delta m_{\PCAC}$ remains constant with $c_A$
within statistical errors.  This analysis suggests a value of $c_A$ 
much larger than the perturbative one, which is not surprising 
given the large coupling.  Unfortunately, a comparison with the quenched case
is not possible since from \cite{Luscher} we have $c_A$ only 
down to $\beta=6.0$ and the $(\beta,\kappa_{\sea})$ parameters we have used
are best compared to a quenched value of $\beta=5.7$, for which the lattice
spacing is almost the same.  A more detailed study, with higher statistics, 
is currently underway.  Future runs with the correct $c_{\SW}$ 
\cite{Jansen} and $b_g$ are in progress.

I would like to thank G.~Martinelli, G.C.~Rossi, C.T.~Sachrajda, S.~Sharpe, 
M.~Testa and all the members of the UKQCD Collaboration for fruitful 
discussions.  We acknowlegde the support of PPARC
through grant GR/L22744 for the time allocation on the Cray T3E.


\begin{thebibliography}{99}    
\bibitem{SW} 
B.~Sheikholeslami and R.~Wholert, \NPB{259} (1985) 572.
\bibitem{Luscher} 
M.~L\"uscher \etal, \NPB{491}(1997)323.
\bibitem{Jansen}
K.~Jansen, R.~Sommer, {\em these proceedings}. 
\bibitem{Sroczynski}
Z.~Sroczynski \etal, {\em these proceedings}.
\bibitem{Guesken}
S.~Guesken, {\em these proceedings}.
\bibitem{Sommer}
R.~Sommer, \NPB{411} (1994) 839.
\bibitem{Allton}
C.R.~Allton \etal, \NPB{489}(1997)427.
\bibitem{Sint}
S.~Sint and R.~Sommer, \NPB{465} (1996) 71.
\bibitem{Martinelli}
G.~Martinelli \etal, hep-lat/9705018.
\bibitem{UKQCD}
UKQCD Collaboration, {\em in preparation}.
\end{thebibliography}
\end{document}